# Ultrafast Coherent Manipulation of Trions in Site-Controlled Nanowire Quantum Dots

K. G. Lagoudakis,[1,*,†] P. L. McMahon,[1,†] C. Dory,[1,†] K. A. Fischer,[1] K. Müller,[1] V. Borish,[1] D. Dalacu,[2] P. J. Poole,[2] M. E. Reimer,[3] V. Zwiller,[4] Y. Yamamoto,[1,5] and J. Vučković[1]

[1]*E.L. Ginzton Laboratory, Stanford University, Stanford, California 94305, USA*
[2] *National Research Council of Canada, Ottawa, K1A 0R6, Canada*
[3] *Institute for Quantum Computing, University of Waterloo, Waterloo, Ontario, N2L 3G1 Canada*
[4] *Department of Applied Physics, Royal Institute of Technology (KTH), 106 91 Stockholm, Sweden*
[5] *National Institute of Informatics, Hitotsubashi 2-1-2, Chiyoda-ku, Tokyo 101-8403, Japan*
*\*Corresponding author: lagous@stanford.edu*
*†These authors contributed equally to this work*

**Physical implementations of large-scale quantum processors based on solid-state platforms benefit from realizations of quantum bits positioned in regular arrays. Self-assembled quantum dots are well-established as promising candidates for quantum optics and quantum information processing, but they are randomly positioned. Site-controlled quantum dots, on the other hand, are grown in pre-defined locations, but have not yet been sufficiently developed to be used as a platform for quantum information processing. In this letter we demonstrate all-optical ultrafast complete coherent control of a qubit formed by the single-spin/trion states of a charged site-controlled nanowire quantum dot. Our results show that site-controlled quantum dots in nanowires are promising hosts of charged-exciton qubits, and that these qubits can be cleanly manipulated in the same fashion as has been demonstrated in randomly-positioned quantum dot samples. Our findings suggest that many of the related excitonic qubit experiments that have been performed over the past 15 years may work well in the more scalable site-controlled systems, making them very promising for the realization of quantum hardware.**

## 1. Introduction

Coherent control of quantum bits (qubits) lies at the heart of quantum computing. Among the wide variety of systems hosting qubits that can be coherently controlled, the platform of self-assembled quantum dots (QDs) is one of the most prominent due to their nanoscale size and the possibility of picosecond-timescale manipulation. Using all-optical techniques, several groups have demonstrated coherent control of excitonic qubits through experiments that involve ultrafast pulses to drive Rabi rotations [1-9] or demonstrate Ramsey interference [10-15]. Similar experiments have been done utilizing biexcitonic states [16-18] and trions [11,19,20]. The self-assembly growth-mechanism however, does not allow for controllable positioning of the qubits and therefore renders such dots imperfect for use in a scalable multi-qubit system. Site-controlled quantum dots have recently emerged as a promising technology in addressing the issue of qubit positioning. Among the existing site-controlled quantum dot technologies [21-27], InAsP quantum dots embedded in deterministically-positioned InP nanowires [28] stand out for their high efficiency [29] single- [30] and entangled-photon [31,32] generation properties. This new quantum dot system gives us an opportunity to revisit the physics of excitonic qubit coherent control in a novel, scalable platform. In this work, we demonstrate complete coherent control of individual spin-trion qubits in site-controlled InAsP nanowire quantum dots under magnetic field by means of resonant multi-pulse excitation. The magnetic field in the Voigt configuration Zeeman-splits the ground and excited states creating a double lambda (Λ) system that we use as our setting for the coherent control experiments.

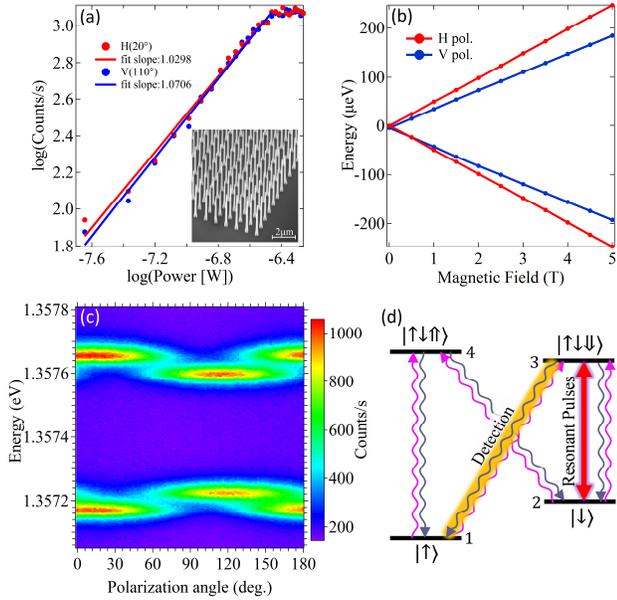

**Fig. 1.** (a) Photoluminescence intensity of the QD emission as a function of the above-band power; the emission is split in two orthogonal linear polarizations. Emissions from both polarizations have a linear power dependence. (b) Magnetic-field dependence of the quantum dot transition energies for the two polarizations, obtained from magneto-photoluminescence spectra (see Supplement 1 for details). (c) Polarization analysis of the photoluminescence when a B=5 T field is applied. (d) Four-level structure of the charged quantum dot in a magnetic field. The optically-excited states are the trions (3,4), whereas the ground states are the spin states (1,2). The grey downward-wavy lines denote spontaneous emission channels. The weak above-band resetting laser is depicted as the violet upward-wavy arrows. Photons are solely detected from the diagonal $|\uparrow\downarrow\Downarrow\rangle\to|\uparrow\rangle$ transition. The driving pulses are resonant with the $|\downarrow\rangle\to|\uparrow\downarrow\Downarrow\rangle$ transition (double sided arrow).

We perform quantum optical modelling that fully captures the observed phenomenology giving rich insights in the underlying physics and the robustness of the system.

## 2. Sample and Experiments

**Nanowire site-controlled QD sample**

The sample we studied has a regular array of nanowire quantum dots similar to the one shown in the inset of Fig. 1(A). The nanowires are grown by vapor-liquid-solid epitaxy on an (111)B InP substrate and deterministic positioning is achieved by masking the sample with a $SiO_2$ mask containing a grid of apertures with a gold nanoparticle centered in each of the apertures. Once the nanowire containing the InAsP QD is grown, it is then surrounded by an InP shell grown in a second step [28].

**Charging and level-structure**

The nanowire QD that we study here is initially characterized with photoluminescence measurements at cryogenic temperatures (T=8.3 K). Excitation of the sample and collection of the emitted photons is done using a 0.5-NA, long-working-distance microscope objective. Above-band excitation yields photoemission from the nanowire QD which is observed in two linear polarizations, and spectral characterization reveals linewidths of ~45±6 µeV. As shown in Fig. 1(a), the emission exhibits clear linear power dependence for both polarizations with saturation occurring at ~420 nW for 780 nm above-band excitation. Determination of the charge state of our quantum dot (negatively charged) is here done by magneto-photoluminescence and spin-pumping measurements in the Voigt configuration (see Supplement 1 and [33]). Application of the magnetic field splits the emission into four distinct spectral lines. The peak locations of these spectral lines and their polarizations are shown in Fig. 1(b). The linear dependence of the splittings on the magnetic field and the opposite polarization between the two inner and outer transitions provides a strong experimental signature of the existence of a charged quantum dot and its characteristic double-Λ system [34,35]. The g-factors for this quantum dot are $g_e$=1.49 and $g_h$=0.22, while the diamagnetic shift factor is 7.13 µeV/T², in good agreement with similar nanowire structures [34]. A complete polarization analysis at maximum magnetic field (B=5 T) is shown in Fig. 1(c). The four-level structure of the system is illustrated in Fig. 1(d), with the two inner and the two outer transitions perpendicularly polarized.

In this paper we define our quantum bit basis as the two levels $|\downarrow\rangle$ and $|\uparrow\downarrow\Downarrow\rangle$ (or levels 2 and 3 shown in Fig. 1(d)). We note that the spin basis used here is along the magnetic field (x-axis). The exact orientation of the nanowire with respect to the magnetic field can be found in Fig. S1 of Supplement 1. We apply a spectrally narrow resonant laser pulse on the $|\downarrow\rangle\to|\uparrow\downarrow\Downarrow\rangle$ transition. The pulse brings the system from the electron spin ground state $|\downarrow\rangle$ to the excited trion state $|\uparrow\downarrow\Downarrow\rangle$, which then radiatively decays via spontaneous emission (downward grey wavy arrows in Fig. 1(d)) either to the $|\uparrow\rangle$ state or to the $|\downarrow\rangle$ state with 50% probability each. If the system decays back to $|\downarrow\rangle$ then the next pulse can re-excite the system to the trion state but if it decays to $|\uparrow\rangle$ the pulse is no longer resonant with the transition unless the system is somehow brought back to the $|\downarrow\rangle$ state. We reset this spin ground state by exciting the system with a weak ~50 nW above-band laser. The net effect of the weak above-band excitation and the spontaneous decay of the trion states to the ground states is that our system is initialized in the state $|\downarrow\rangle$ with 50% probability. We perform a measurement of the qubit state by counting photons emitted by the diagonal transition $|\uparrow\downarrow\Downarrow\rangle\to|\uparrow\rangle$ with a free-space single-photon-counting module (SPCM); photons from other decay pathways are excluded from detection by spectral filtering using a custom made double monochromator of 1.75m overall length.

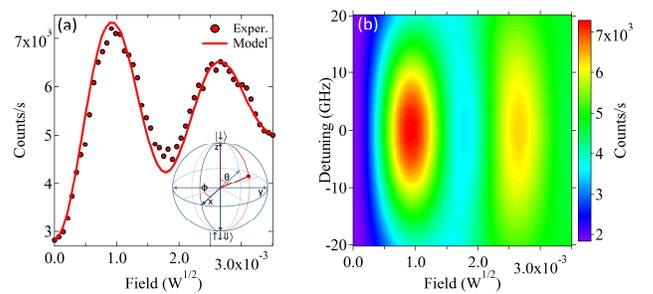

**Fig. 2.** (a) Rabi oscillations of the trion qubit. The detected counts from the $|\uparrow\downarrow\Downarrow\rangle\to|\uparrow\rangle$ transition are proportional to the probability of the qubit being in state $|\uparrow\downarrow\Downarrow\rangle$. The pulses, which are resonant with the $|\downarrow\rangle\to|\uparrow\downarrow\Downarrow\rangle$ transition cause rotations of the qubit, which begins in the state $|\downarrow\rangle$. The solid red line is a fit from the model. The inset depicts the Bloch sphere and its principle axes. (b) Modelled Rabi oscillations for a range of driving pulse detunings.

**Rabi oscillations**

To demonstrate coherent control of the trion qubit we drive the $|\downarrow\rangle \rightarrow |\uparrow\downarrow\Downarrow\rangle$ transition with ~20 ps pulses of variable amplitude that we prepare using a pulse shaper. These pulses are derived from a Ti:Sapphire picosecond pulsed laser with 80.2-MHz repetition rate. On the Bloch sphere, individual optical pulses rotate the Bloch vector about the x-axis by an angle $\theta$ proportional to the area of the pulse. In Fig. 2(a) we provide the photon counts measured from the diagonal transition $|\uparrow\downarrow\Downarrow\rangle \rightarrow |\uparrow\rangle$ as a function of the pulse area. We observe clear Rabi oscillations that we can trace from 0 all the way to approximately $4\pi$. The damping of the oscillations is likely due to excitation-related dephasing [36], phonon relaxation, and spontaneous emission. We fit the experimentally observed oscillations using a model implemented with the Quantum Optics Toolbox in Python (QuTiP) [37], which is described in detail in the Modelling section. Using the parameters that yielded the best fit to the experimental data, we simulated the effect of the detuning of the resonant driving field to gain better insight in the robustness of the process. A detuned pulse will drive the Bloch vector about an axis rotated by $\varphi \propto \Delta\omega_L$ with respect to the x-axis of the Bloch sphere, as depicted in the inset of Fig. 2(a). In Fig. 2(b) the modeled Rabi oscillations are presented as a function of the resonant pulse detuning and driving power. The oscillations persist for detunings of up to ±20 GHz with a strong reduction in their amplitude as the laser is tuned out of resonance.

**Ramsey interference**

As we mentioned previously, an individual resonant pulse rotates the Bloch vector about the x-axis. Using a second pulse applied after a delay causes the qubit to undergo a rotation about a second axis that is at an angle $\varphi = \omega_L \cdot \Delta t$ with respect to the x-axis. A Mach-Zehnder interferometer, with a delay stage in one arm for coarse delay tuning $t_c$ and a piezo-controlled mirror in the other arm for fine delay tuning $t_f$, splits the initial pulses into two copies with variable interpulse delay. This dual-pulse train is used for the Ramsey interference experiments. In such an experiment, the pulse areas are chosen so that each individual pulse causes a rotation by $\pi/2$ rad. The first pulse rotates the Bloch vector about the x-axis by $\pi/2$ rad, creating a coherent superposition of ground $|\downarrow\rangle$ and excited $|\uparrow\downarrow\Downarrow\rangle$ states with equal amplitudes.

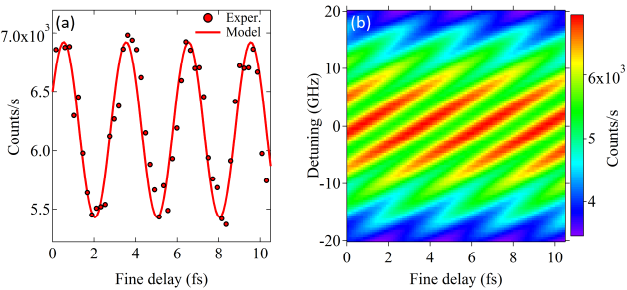

**Fig. 3.** (a) Ramsey interference experiment where the qubit is driven by two $\pi/2$ pulses separated by a variable delay. Here, the coarse delay is 80 ps and the fine delay is scanned over 11 fsec revealing oscillations that are due to quantum interference. The solid red line is a fit from the model. (b) Modelled Ramsey interference for a range of resonant pulse detunings. Detuning the pulse introduces a linear phase shift of the interference fringes and causes a reduction in their amplitude.

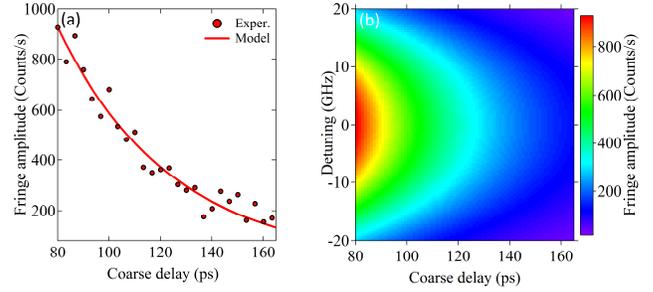

**Fig. 4.** (a) Determination of $T_2^*$ using the decay of the Ramsey fringe amplitude as a function of the coarse delay. Note that the initial delay that is used in this experiment is 80 ps. The solid red line is a fit from the model with a decay time of 43 ps. (b) Modelled Ramsey fringe amplitude decay as a function of the coarse delay for a range of resonant pulse detunings. Although the amplitude gradually reduces when the detuning is increased, the decay time remains the same.

If the delay between the pulses is such that the second rotation axis is at an angle $\varphi = 2n\pi$ (for $n$ integer) with respect to the x-axis, then the second pulse will rotate the Bloch vector to the excited state $|\uparrow\downarrow\Downarrow\rangle$, resulting in maximal detected counts. If on the other hand the delay results in the second rotation axis being at an angle $\varphi = (2n+1)\pi$ with respect to the x-axis, then the second pulse will bring the Bloch vector back to the ground state $|\downarrow\rangle$, giving a minimum in detected counts. Recording the detected counts for a range of interpulse delays allows one to observe Ramsey interference fringes. Fig. 3(a) shows the experimentally observed oscillations as a function of the piezo-controlled fine interpulse delay over the range $\Delta t_f \in (0,11)$ fs. The two $\pi/2$ pulses additionally have a coarse delay of $\Delta t_c = 80$ ps, which eliminates any optical interference between the pulses themselves so that the observed oscillations only come from the Ramsey interference. The solid line in Fig. 3(a) is a fit from the model that we further use to demonstrate the effect of the pulse frequency detuning in Fig. 3(b).

As shown by the model, increasing detunings lead to a clear Ramsey interference amplitude reduction while the phase of the fringes shows a linear dependence on the pulse detuning $\Delta\omega_L$. The phase shift is a consequence of the pulse detuning; the detuning causes the rotation axis to be shifted by an angle $\varphi = (\omega_0 + \Delta\omega_L) \cdot (t_c + t_f)$ with respect to the non-detuned-pulse axis of rotation. In principle this effect can be used to perform a Ramsey interference experiment by keeping the interpulse delay fixed and just varying the pulse detuning [14].

**Determination of coherence time**

The decay of the Ramsey interference amplitude for longer delays provides a measurement of the extrinsic dephasing time $T_2^*$ [12]. To access information on the decay of the Ramsey interference we record the amplitude of the oscillations for a range of coarse delays. Starting with the initial delay of $\Delta t_c = 80$ ps, we gradually increase the delay to 180 ps in steps of ~3.34 ps. For each of the coarse delays we repeat the Ramsey interference experiment by finely scanning the interpulse delay over the range $\Delta t_f \in (0,11)$ fsec using the piezo-controlled delay and recording the signal amplitude.

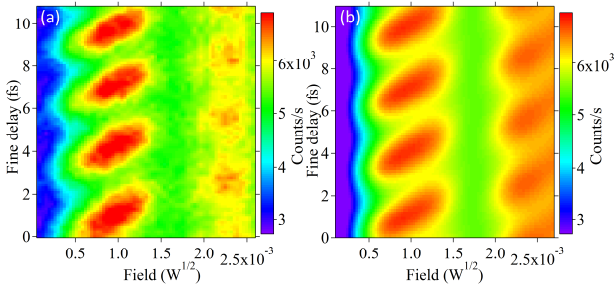

**Fig. 5.** (a) Detected counts when the system is manipulated with two pulses of variable power and variable delay. This allows access to the full Bloch sphere (complete coherent control). The tilt of the lobes for low pulse powers originates from the non-zero detuning of the resonant pulses used in this experiment ($\Delta\omega_L/2\pi \sim 14.5\,\text{GHz}$). (b) Modelled complete coherent control, with parameters set to match those used in the experiment.

---

In Fig. 4(a) we provide the measured amplitude of the Ramsey interference oscillations for all the coarse delays. The solid line is a fit from our model and the decay time corresponds to $T_2^* = 43$ ps which is much shorter than the lifetime of the trion itself ~1 ns [30-32] and in good agreement with previously reported values for trions in other quantum dot systems [20]. Using our quantum optical model we investigated the effect of the pulse detuning and found that although it affects the overall amplitude of the oscillations, the decay time remains unchanged for the complete range of pulse detunings investigated, and is mostly affected by the intrinsic and phonon-induced dephasing [36]. Fig. 4(b) shows the dependence of the interference amplitude as a function of the pulse detuning, highlighting the reduction in interference amplitude for increasing detuning. Experiments performed on other nearby quantum dots have yielded similar decay times.

**Complete coherent control experiment**

In order to demonstrate universal single-qubit gate operation with our site-controlled nanowire quantum dot qubit, we performed a variant of the standard Ramsey experiment.

In addition to varying the delay between the two rotation pulses, we also vary their power (i.e., area). Doing so provides access to states that are anywhere on the Bloch sphere (complete coherent control). This can be achieved because all locations on the Bloch sphere can be reached by performing two rotation operations of the Bloch vector if both the angle of rotation and the axis of the second rotation relative to the first are controlled. Fig. 5(a) shows the detected counts as a function of the rotation pulse power and the interpulse delay. In this experiment the pulses are slightly detuned from resonance by $\Delta\omega_L/2\pi \sim 14.5$ GHz. This is the origin of the slightly tilted lobe structure in Fig. 5(a), which we reproduced using our quantum optical model, as shown in Fig. 5(b). For simulated results from other pulse detunings the reader can refer to Fig. S3 of the Supplement 1.

**Modelling**

The simulations presented in our work were performed using the Quantum Toolbox in Python (QuTiP) [37].

The four-level system, driven by a pump laser with frequency

$$\omega_L = \omega_0 - \frac{\Delta E_{tr}}{2} - \frac{\Delta E_{gs}}{2} - \Delta\omega_L,$$

where $\Delta\omega_L$ denotes the detuning of the laser, is described by the Hamiltonian $\mathcal{H} = \mathcal{H}_o + \mathcal{H}_D$, where

$$\mathcal{H}_o = -\frac{\Delta E_{gs}}{2}s_{11} + \frac{\Delta E_{gs}}{2}s_{22} + \left(\omega_0 - \frac{\Delta E_{tr}}{2}\right)s_{33} + \left(\omega_0 + \frac{\Delta E_{tr}}{2}\right)s_{44}$$

represents the unperturbed QD system, where $\omega_0$ is the zero field splitting and $\Delta E_{gs}$, $\Delta E_{tr}$ are the ground and trion state splittings in the presence of a magnetic field. $s_{ii}$ are the projection operators of the self-energy terms. The driving term is:

$$\mathcal{H}_D = (s_{14} + s_{23})\left(\Omega(t) + \Omega(t-\Delta t)e^{i\omega_L \Delta t}\right) + \\ (s_{41} + s_{32})\left(\Omega(t) + \Omega(t-\Delta t)e^{-i\omega_L \Delta t}\right)$$

in which the driving strength $\Omega$ is proportional to the driving electric field magnitude and $\Delta t = t_c + t_f$ is the total delay time between the first and the second excitation pulse. Due to the selection rules governing the system, only transitions $s_{41}$, $s_{14}$ and $s_{32}$, $s_{23}$ are driven by the laser. Since the detected signal comes from the spontaneous emission collected from the transition $s_{31}$, the population level at time $t_1 = t_0 + \Delta t + \tau_{\text{FWHM}}$ ($t_0$ is the time of arrival of the first pulse) is calculated by integrating the master equation

$$\frac{d\tilde{\rho}(t)}{dt} = -i\left[\tilde{\mathcal{H}}, \tilde{\rho}(t)\right] + \sum_j \mathcal{L}(c_j),$$

With $\mathcal{L}(c_j)$ the Lindblad superoperators of the collapse operators $c_j$. The effect of the above-band laser is modelled as the inverse of spontaneous emission. Our simulations of the final density matrix provide the photon count rates expected from each measurement (up to a normalization factor). The complete set of parameters used for the simulations that we performed here, is in reference [38].

## 3. Summary

In this work we have demonstrated complete coherent control of a trion-based qubit in a site-controlled nanowire quantum dot. We used the double Λ structure of the charged quantum dot in a high magnetic field as a means to spectrally separate the rotation pulses from the detection channel and using ultrafast pulse sequences we showed Rabi oscillations, Ramsey interference and complete coherent control of our qubit.

The short coherence time of the trion qubit is here a limiting factor, but newer generation samples with enhanced growth conditions hold great potential for coherence time improvements [29]. Moreover the recent development of both electrical [39] and strain tuning [40] of these nanowire QDs could help alleviate the inhomogeneous broadening making this platform very promising for quantum hardware and opening the path for more robust qubits [41-42] to be implemented in site-controlled nanowire quantum dots. Nanowire-based qubits do not admit a direct mechanism for coupling neighboring (nor distant) qubits, but by adapting schemes developed for scalable trapped-ion quantum processors [43], in which distant qubits are entangled optically, one can imagine an architecture for a quantum processor based on nanowire-QD qubits. Although the replication of long-coherence-time qubit experiments in a site-controlled quantum dot platform is a major challenge in the QD roadmap for building a quantum repeater [44], technological advancements in several promising site-controlled QD platforms [21-27] make this prospect appear within our grasp.

**Funding.** We acknowledge support from the Army Research Office (grant W911NF1310309) and the National Science Foundation (grant 1503759). We also acknowledge support from The Cabinet Office, Government of Japan, and Japan Society for the Promotion of Science (JSPS) through the Funding Program for World-Leading Innovative R&D on Science and Technology (FIRST Program). KGL acknowledges support by the Swiss National Science Foundation. PLM was supported by a Stanford Nano- and Quantum Science and Engineering Postdoctoral Fellowship. KAF acknowledges support from the Lu Stanford Graduate Fellowship and the National Defense Science and Engineering Fellowship. KM acknowledges support from the Alexander von Humboldt Foundation. VB acknowledges support from the National Science Foundation Graduate Research Fellowship.

See Supplement 1 for supporting content.

## REFERENCES

1. N. H. Bonadeo, J. Erland, D. Gammon, D. Park, D. S. Katzer, and D. G. Steel, Coherent Optical Control of the Quantum State of a Single Quantum Dot. *Science*. **282**, 1473–1476 (1998).
2. A. Zrenner, E. Beham, S. Stufler, F. Findeis, M. Bichler, and G. Abstreiter, Coherent properties of a two-level system based on a quantum-dot photodiode. *Nature*. **418**, 612–614 (2002).
3. T. H. Stievater, X. Li, D. G. Steel, D. Gammon, D. S. Katzer, D. Park, C. Piermarocchi, and L. J. Sham, Rabi Oscillations of Excitons in Single Quantum Dots. *Phys. Rev. Lett.* **87**, 133603 (2001).
4. B. Patton, U. Woggon, W. Langbein, Coherent Control and Polarization Readout of Individual Excitonic States. *Phys. Rev. Lett.* **95**, 266401 (2005).
5. Q. Q. Wang, A. Muller, P. Bianucci, E. Rossi, Q. K. Xue, T. Takagahara, C. Piermarocchi, A. H. MacDonald, and C. K. Shih, Decoherence processes during optical manipulation of excitonic qubits in semiconductor quantum dots. *Phys. Rev. B*. **72**, 035306 (2005).
6. S. Stufler, P. Machnikowski, P. Ester, M. Bichler, V. M. Axt, T. Kuhn, and A. Zrenner, Two-photon Rabi oscillations in a single $In_xGa_{1-x}As$-GaAs quantum dot. *Phys. Rev. B*. **73**, 125304 (2006).
7. A. J. Ramsay, R. S. Kolodka, F. Bello, P. W. Fry, W. K. Ng, A. Tahraoui, H. Y. Liu, M. Hopkinson, D. M. Whittaker, A. M. Fox, and M. S. Skolnick, Coherent response of a quantum dot exciton driven by a rectangular spectrum optical pulse. *Phys. Rev. B*. **75**, 113302 (2007).
8. R. Melet, V. Voliotis, A. Enderlin, D. Roditchev, X. L. Wang, T. Guillet, and R. Grousson, Resonant excitonic emission of a single quantum dot in the Rabi regime. *Phys. Rev. B*. **78**, 073301 (2008).
9. H. Takagi, T. Nakaoka, K. Watanabe, N. Kumagai, Y. Arakawa, Coherently driven semiconductor quantum dot at a telecommunication wavelength. *Optics Express*. **16**, 13949 (2008).
10. H. Htoon, T. Takagahara, D. Kulik, O. Baklenov, A. L. Holmes, and C. K. Shih, Interplay of Rabi Oscillations and Quantum Interference in Semiconductor Quantum Dots. *Phys. Rev. Lett.* **88**, 087401 (2002).
11. L. Besombes, J. J. Baumberg, J. Motohisa, Coherent Spectroscopy of Optically Gated Charged Single InGaAs Quantum Dots. *Phys. Rev. Lett.* **90**, 257402 (2003).
12. R. S. Kolodka, A. J. Ramsay, J. Skiba-Szymanska, P. W. Fry, H. Y. Liu, A. M. Fox, and M. S. Skolnick, Inversion recovery of single quantum-dot exciton based qubit. *Phys. Rev. B*. **75**, 193306 (2007).
13. S. Stufler, P. Ester, A. Zrenner, M. Bichler, Quantum optical properties of a single $In_xGa_{1-x}As$-GaAs quantum dot two-level system. *Phys. Rev. B*. **72**, 121301 (2005).
14. S. Stufler, P. Ester, A. Zrenner, M. Bichler, Ramsey Fringes in an Electric-Field-Tunable Quantum Dot System. *Phys. Rev. Lett.* **96**, 037402 (2006).
15. K. Müller, T. Kaldewey, R. Ripszam, J. S. Wildmann, A. Bechtold, M. Bichler, G. Koblmüller, G. Abstreiter, and J. J. Finley All optical quantum control of a spin-quantum state and ultrafast transduction into an electric current. *Scientific Reports*. **3**, 1906 (2013).
16. Y. Kodriano, I. Schwartz, E. Poem, Y. Benny, R. Presman, T. A. Truong, P. M. Petroff, and D. Gershoni, Complete control of a matter qubit using a single picosecond laser pulse. *Phys. Rev. B*. **85**, 241304 (2012).
17. E. Poem, O. Kenneth, Y. Kodriano, Y. Benny, S. Khatsevich, J. E. Avron, and D. Gershoni, Optically Induced Rotation of an Exciton Spin in a Semiconductor Quantum Dot. *Phys. Rev. Lett.* **107**, 087401 (2011).
18. Y. Benny, S. Khatsevich, Y. Kodriano, E. Poem, R. Presman, D. Galushko, P. M. Petroff, and D. Gershoni, Coherent Optical Writing and Reading of the Exciton Spin State in Single Quantum Dots. *Phys. Rev. Lett.* **106**, 040504 (2011).
19. J. S. Pelc, L. Yu, K. De Greve, P. L. McMahon, C. M. Natarajan, V. Esfandyarpour, S. Maier, C. Schneider, M. Kamp, S. Höfling, R. H. Hadfield, A. Forchel, Y. Yamamoto, and M. M. Fejer, Downconversion quantum interface for a single quantum dot spin and 1550-nm single-photon channel. *Optics Express*. **20**, 27510 (2012).
20. M. V. Gurudev Dutt, J. Cheng, Y. Wu, X. Xu, D. G. Steel, A. S. Bracker, D. Gammon, S. E. Economou, R.-B. Liu, and L. J. Sham, Ultrafast optical control of electron spin coherence in charged GaAs quantum dots. *Phys. Rev. B*. **74**, 125306 (2006).
21. A. Jamil, J. Skiba-Szymanska, S. Kalliakos, A. Schwagmann, M. B. Ward, Y. Brody, D. J. P. Ellis, I. Farrer, J. P. Griffiths, G. A. C. Jones, D. A. Ritchie, and A. J. Shields, On-chip generation and guiding of quantum light from a site-controlled quantum dot. *Appl. Phys. Lett.* **104**, 101108 (2014).
22. K. D. Jöns, P. Atkinson, M. Müller, M. Heldmaier, S. M. Ulrich, O. G. Schmidt, and P. Michler, Triggered Indistinguishable Single Photons with Narrow Line Widths from Site-Controlled Quantum Dots. *Nano Lett.* **13**, 126–130 (2013).
23. C. Schneider, T. Heindel, A. Huggenberger, T. A. Niederstrasser, S. Reitzenstein, A. Forchel, S. Höfling, and M. Kamp, Microcavity enhanced single photon emission from an electrically driven site-controlled quantum dot. *Appl. Phys. Lett.* **100**, 091108 (2012).
24. J. Tommila, V. V. Belykh, T. V. Hakkarainen, E. Heinonen, N. N. Sibeldin, A. Schramm, and M. Guina, Cavity-enhanced single photon emission from site-controlled In(Ga)As quantum dots fabricated using nanoimprint lithography. *Appl. Phys. Lett.* **104**, 213104 (2014).
25. M. K. Yakes, L. Yang, A. S. Bracker, T. M. Sweeney, P. G. Brereton, M. Kim, C. S. Kim, P. M. Vora, D. Park, S. G. Carter, and D. Gammon, Leveraging Crystal Anisotropy for Deterministic Growth of InAs Quantum Dots with Narrow Optical Linewidths. *Nano Lett.* **13**, 4870–4875 (2013).
26. M. J. Holmes, K. Choi, S. Kako, M. Arita, Y. Arakawa, Room-Temperature Triggered Single Photon Emission from a III-Nitride Site-Controlled Nanowire Quantum Dot. *Nano Lett.* **14**, 982–986 (2014).
27. G. Juska, V. Dimastrodonato, L. O. Mereni, A. Gocalinska, and E. Pelucchi, Towards quantum-dot arrays of entangled photon emitters. *Nat Photon.* **7**, 527–531, (2013).
28. D. Dalacu, K. Mnaymneh, J. Lapointe, X. Wu, P. J. Poole, G. Bulgarini, V. Zwiller, and M. E. Reimer, Ultraclean Emission from InAsP Quantum Dots in Defect-Free Wurtzite InP Nanowires. *Nano Lett.* **12**, 5919–5923 (2012).
29. M. E. Reimer, G. Bulgarini, A. Fognini, R. W. Heeres, B. J. Witek, M. A. M. Versteegh, A. Rubino, T. Braun, M. Kamp, S. Höfling, D. Dalacu, J. Lapointe, P. J. Poole, and V. Zwiller, Overcoming power broadening of the quantum dot emission in a pure wurtzite nanowire. *Phys. Rev. B* **93**, 195316 (2016).
30. M. E. Reimer, G. Bulgarini, N. Akopian, M. Hocevar, M. B. Bavinck, M. A. Verheijen, E. P. A. M. Bakkers, L. P. Kouwenhoven, and V. Zwiller, Bright single-photon sources in bottom-up tailored nanowires. *Nat Commun.* **3**, 737 (2012).
31. T. Huber, A. Predojević, M. Khoshnegar, D. Dalacu, P. J. Poole, H. Majedi, and G. Weihs, Polarization Entangled Photons from Quantum Dots Embedded in Nanowires. *Nano Lett.* **14,** 7107–7114 (2014).
32. M. A. M. Versteegh, M. E. Reimer, K. D. Jöns, D. Dalacu, P. J. Poole, A. Gulinatti, A. Giudice, and V. Zwiller, Observation of strongly entangled photon pairs from a nanowire quantum dot. *Nat Commun.* **5**, 5298 (2014).
33. K. G. Lagoudakis, P. L. McMahon, K. A. Fischer, S. Puri, K. Müller, D. Dalacu, P. J. Poole, M. E. Reimer, V. Zwiller, Y. Yamamoto and J. Vučković Initialization of a spin qubit in a site-controlled nanowire quantum dot. *New J. Phys.* **18**, 053024 (2016).
34. B. J. Witek, R. W. Heeres, U. Perinetti, E. P. A. M. Bakkers, L. P. Kouwenhoven, and V. Zwiller, Measurement of the g-factor tensor in a quantum dot and disentanglement of exciton spins. *Phys. Rev. B* **84**, 195305 (2011).
35. M. Bayer, G. Ortner, O. Stern, A. Kuther, A. A. Gorbunov, A. Forchel, P. Hawrylak, S. Fafard, K. Hinzer, T. L. Reinecke, S. N. Walck, J. P. Reithmaier, F.

Klopf, and F. Schäfer, Fine structure of neutral and charged excitons in self-assembled In(Ga)As/(Al)GaAs quantum dots. *Phys. Rev. B*. **65**, 195315 (2002).

36. A. J. Ramsay, T. M. Godden, S. J. Boyle, E. M. Gauger, A. Nazir, B. W. Lovett, A. M. Fox, and M. S. Skolnick, Phonon-Induced Rabi-Frequency Renormalization of Optically Driven Single InGaAs/GaAs Quantum Dots. *Phys. Rev. Lett.* **105**, 177402 (2010).

37. J. R. Johansson, P. D. Nation, F. Nori, QuTiP 2: A Python framework for the dynamics of open quantum systems. *Comput. Phys. Commun*. **184**, 1234–1240 (2013).

38. The exact simulation parameters that were used are:
$\Delta E_{gs}/2\pi = 104.2\,\text{GHz}$, $\Delta E_{tr}/2\pi = 15.1\,\text{GHz}$, $\omega_0/2\pi = 333 \cdot 10^3\,\text{GHz}$,
$\Omega/2\pi \in (0, 4.46)\,\text{GHz}$, $\Delta E_{tr}/2\pi = 15.1\,\text{GHz}$, $\Delta\omega_L/2\pi \in (-20, 20)\,\text{GHz}$,
$\gamma_{\text{spontaneous}} = 1\,\text{ns}^{-1}$, $\gamma_{\text{pumping}} = \frac{50}{420} \cdot \gamma_{\text{spontaneous}}$, $\gamma_{\text{dephasing}} = \frac{1}{145 \cdot 10^{-3}}\,\text{ns}^{-1}$,
$\alpha_{\text{phonon}} = \frac{1}{3.6 \cdot 10^{-3}}\,\text{ns}^{-1}$, $\tau_{\text{FWHM}} = 23\,\text{psec}$ and the time dependent collapse operators are $s_{jj}\sqrt{\alpha_{\text{phonon}}}\left(\Omega(t) + \Omega(t - \Delta t)\right)$, $j = 3, 4$.

39. M. E. Reimer, M. P. van Kouwen, A. W. Hidma, M. H. M. van Weert, E. P. A. M. Bakkers, L. P. Kouwenhoven, and V. Zwiller, Electric Field Induced Removal of the Biexciton Binding Energy in a Single Quantum Dot. Nano Lett. 11, 645–650 (2011).

40. Y. Chen, I. E. Zadeh, K. D. Jöns, A. Fognini, M. E. Reimer, J. Zhang, D. Dalacu, P. J. Poole, F. Ding, V. Zwiller, and O. G. Schmidt, Controlling the exciton energy of a nanowire quantum dot by strain fields, Appl. Phys. Lett. 108, 182103 (2016).

41. A. J. Ramsay, A review of the coherent optical control of the exciton and spin states of semiconductor quantum dots. *Semiconductor Science and Technology*. **25**, 103001 (2010).

42. K. D. Greve, D. Press, P. L. McMahon, Y. Yamamoto, Ultrafast optical control of individual quantum dot spin qubits. *Reports on Progress in Physics*. **76**, 092501 (2013).

43. C. Monroe, J. Kim, Scaling the Ion Trap Quantum Processor. *Science*. **339**, 1164-1169 (2013).

44. P. L. McMahon, K. De Greve, in *Engineering the Atom-Photon Interaction*, A. Predojević, M. W. Mitchell, Eds. (Springer, London, 2015), Chap. 14.

# Ultrafast Coherent Manipulation of Trions in Site-Controlled Nanowire Quantum Dots: Supplementary Material

K. G. Lagoudakis,[1,*,†] P. L. McMahon,[1,†] C. Dory,[1,†] K. A. Fischer,[1] K. Müller,[1] V. Borish,[1] D. Dalacu,[2] P. J. Poole,[2] M. E. Reimer,[3] V. Zwiller,[4] Y. Yamamoto,[1,5] and J. Vučković[1]

[1] E.L. Ginzton Laboratory, Stanford University, Stanford, California 94305, USA
[2] National Research Council of Canada, Ottawa, K1A 0R6, Canada
[3] Institute for Quantum Computing, University of Waterloo, Waterloo, Ontario, N2L 3G1 Canada
[4] Department of Applied Physics, Royal Institute of Technology (KTH), 106 91 Stockholm, Sweden
[5] National Institute of Informatics, Hitotsubashi 2-1-2, Chiyoda-ku, Tokyo 101-8403, Japan
*Corresponding author: lagous@stanford.edu
†These authors contributed equally to this work

This document provides supplementary information to the article "Ultrafast Coherent Manipulation of Trions in Site-Controlled Nanowire Quantum Dots". In particular we are providing details about the experimental setup used for the magnetic spectroscopy and coherent control of the trions in InAsP QDs embedded in InP nanowires. We show the detailed magnetic spectroscopy analysis that includes the raw spectra, the fits and the extracted diamagnetic shift factor. Finally, to complement the modelling of the main manuscript we provide the complete coherent control modelled results for two additional resonant pulse detunings.

**Experimental setup.** The experimental setup used for the trion coherent control experiments is shown in Fig. S1. An above-band diode laser is used for resetting the spin state, and a Ti:Sapphire picosecond pulsed laser provides the pulses for the coherent control. A Mach-Zehnder interferometer splits the pulses in two copies in the two interferometer arms. The ~3-psec pulses are shaped to ~20 psec at the "pulse filtering" stage and combined with the weak above-band laser on a beamsplitter before being focused on the nanowire quantum dot as shown in Fig. S1(a). A polarizer in the detection path allows for polarization-resolved measurements. The signal, once passed through the polarizer, is sent to a custom-made double monochromator with a ~10μeV resolution, which resolves all the spectral lines of the system. Using the output slit, the optical transition of interest is sent to a free-space SPCM that detects the emitted photons. Fig. S1(b) depicts the experimental scheme and highlights the orientation of the main polarization axes with respect to the magnetic field orientation.

**Magnetic spectroscopy.** We performed a complete characterization of the nanowire QD as a function of the magnetic field using a superconducting magnet. The quantum dot was initially characterized at zero field to extract spectroscopic information about its optical emission linewidth; a fit to the spectrum yielded a 45-μeV linewidth, as shown in Fig. S2(a). Gradual application of the magnetic field clearly splits the spectral lines of the nanowire quantum dot into a quadruplet, which indicates that the dot is charged. Fig. S2(b) shows the evolution of the spectra as a function of the magnetic field for the two main polarizations. Multipeak fitting allows extraction of the peak locations of all four spectral lines as a function of the magnetic field. The locations of the peaks are shown in Fig. S2(c). The linear Zeeman shifts are present along with the quadratic diamagnetic shift; the diamagnetic shift moves all four spectral lines higher in energy.

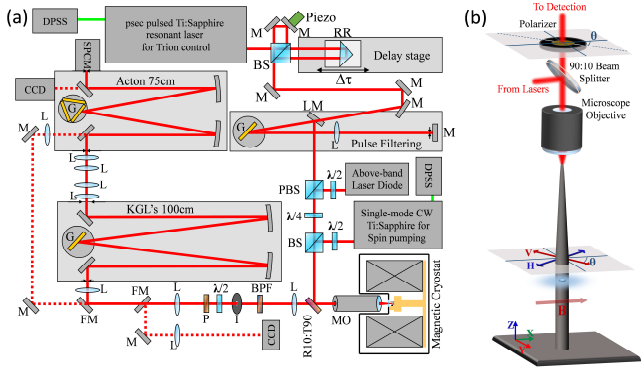

**Fig. S1.** Complete coherent control setup and detection scheme. (a) Experimental setup for the coherent control of trions. DPSS: diode pumped solid state laser, SPCM: single photon counting module, CCD: charge coupled device, M: mirror, FM: flip mirror, LM: lowered mirror, L: lens, P: polarizer, BPF: bandpass filter, I: iris, MO: microscope objective, BS: beam splitter, G: grating, R10:T90: 90-10 partially reflective beam splitter, λ/2: half-wave plate, RR: retroreflector. (b) Schematic representation of optical excitation and detection scheme. The main polarization axes are defined with respect to the orientation of the magnetic field. The quantum dot is depicted as the blue oval close to the center of the nanowire.

The diamagnetic shift is extracted by calculating the midpoint energy between the highest and lowest energy peaks of the same polarization and is shown in Fig. S2(d).

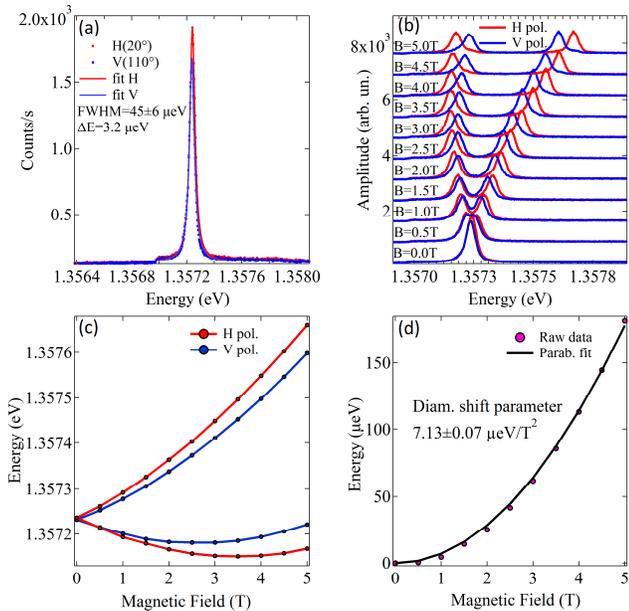

**Fig. S2.** Magnetic spectroscopy of the nanowire quantum dots. (a) Photoluminescence measurement of the quantum dot for above-band excitation for the two main polarizations. The spectral lines are fitted with Lorentzians of 45 ± 6 μeV FWHM and the energy splitting between of the two polarizations is only ΔE=3.2 ± 1 μeV. (b) Magnetic spectroscopy of the nanowire dot for increasing magnetic field in the Voigt configuration under above-band excitation. The spectral structure quickly develops into a quadruplet. Note that the spectra for increasing field have been displaced vertically to help the reader visualize the evolution of the spectra. (c) Fitted locations of the individual spectral peaks. (d) Diamagnetic shift extracted from the peak locations of (c).

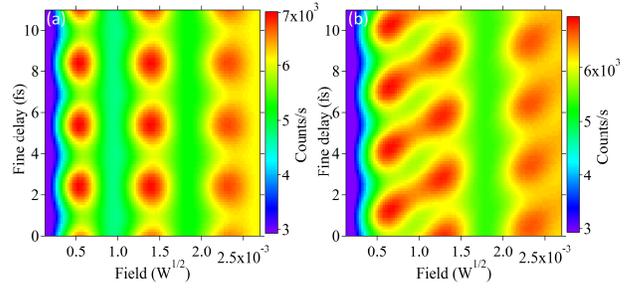

**Fig. S3.** Modelling of complete coherent control experiment for two different detunings. (a) Complete coherent control experiment for zero detuning of the resonant pulses. (b) Here, the detuning was set to $\Delta\omega_L/2\pi = 9.55\,\text{GHz}$. The individual lobes at fields 0.55 and 1.4 W$^{1/2}$ progressively merge with each other for increasing detuning.

The solid line is a quadratic fit, from which we found the diamagnetic shift factor to be 7.13±0.07 μeV/T$^2$. Subtraction of the diamagnetic shift from the raw peak locations of Fig. S2(c) reveals the purely linear Zeeman splitting that is shown in Fig. 1(b) of the manuscript. The $g$-factors that we extracted here are 0.22 and 1.49. A common trait of InAsP QDs in InP nanowires is the large mismatch of the electron and hole $g$-factors with $g_h$ being always much smaller, similar to quantum wells. This allows us to identify that $g_e$=1.49 and $g_h$=0.22. For more information on the g-factors as well as typical values of exchange interactions in InP nanowire QDs the reader can refer to the work by Witek et al. [34]. For the quantum dot investigated in this work, when we drive the lowest energy outer transition, we observe photons from the inner highest energy transition. This suggests that the lambda system we are investigating has the level structure depicted in Fig. 1(d) of the main manuscript (or the insets of Fig. S4) with the ground states being the ones having the largest splitting. We can therefore safely attribute the ground state splitting to the electron $g$-factor and the trion state splitting to the hole $g$-factor indicating that we have a negatively charged quantum dot.

**Complete coherent control for various detunings.** Dual pulse excitation with varying amplitudes and varying interpulse delays allows access to the full Bloch sphere. In the manuscript we have provided the modelled complete coherent control experiment for the detuning used in the actual experiment ($\Delta\omega_L/2\pi \sim 14.5\,\text{GHz}$). To further illustrate the effect of the resonant pulse detuning, we provide the same modelled experiment of complete coherent control for two more detunings. In particular we calculated the behavior of the system for the on-resonance case, with $\Delta\omega_L/2\pi = 0\,\text{GHz}$ as shown in Fig. S3(a) and for $\Delta\omega_L/2\pi = 9.55\,\text{GHz}$ as shown in Fig. S3(b). Increasing the detuning results in a gradual merging of the individual lobes because of the additional angle imposed on the axis of rotation of the Bloch vector. Further increasing the rotation pulse detuning leads to complete merging of the lobes as shown in Fig. 5(b) of the manuscript.

**Spin-Pumping experiments.** We also performed spin pumping experiments on the nanowire QD investigated in this work as shown in Fig. S4(a)-(d). We performed spin pumping in all four possible configurations to further establish the double-Λ structure of our negatively charged QD in the Voigt Magnetic field. Spin pumping is here performed using a monomode tunable CW laser ($P_{pump}$ = 110nW) with which we resonantly drive the spin-ground-state to trion transitions in combination with a weak above-band laser ($P_{rand}$ = 10nW) to randomize the population of the ground spin states. For these experiments each Λ system is pumped along one transition while we only detect photons on the other (orthogonally polarized) transition of the pumped Λ system (see insets of Figs 4(a)-(d)). When the frequency of the CW laser is scanned across one trion transition (for example vertical), while recording photons from the other transition (for example diagonal), a Lorentzian peak can be seen as shown in Fig. S4 (a). Once the randomization laser is turned off, the counts drop to the background detector counts even though the resonant laser is still being scanned across the transition. This indicates high fidelity spin initialization in our nanowire QD. Fig. S4 (a) shows spin pumping when the system is pumped at the highest vertical energy transition 1-4 while detection is along the diagonal transition 4-2. Fig S4 (b) shows spin pumping when the system is driven on 1-3 while detected photons come from 3-2. Fig. S4 (c) shows spin pumping while driving transition 2-4 and detecting transition 4-1 and finally Fig. S4 (d) shows spin pumping along transition 2-3 for photons detected on transition 3-1. In all four insets, the randomization laser is depicted as the upward violet wavy arrows whereas spontaneous emission is depicted as the downward grey wavy arrows. The resonant CW laser is shown as a solid upward arrow and the spontaneous decay detection channel is highlighted in yellow. For a more in-depth investigation of spin pumping in deterministically positioned nanowire QDs like the one studied here, the reader can refer to reference [2].

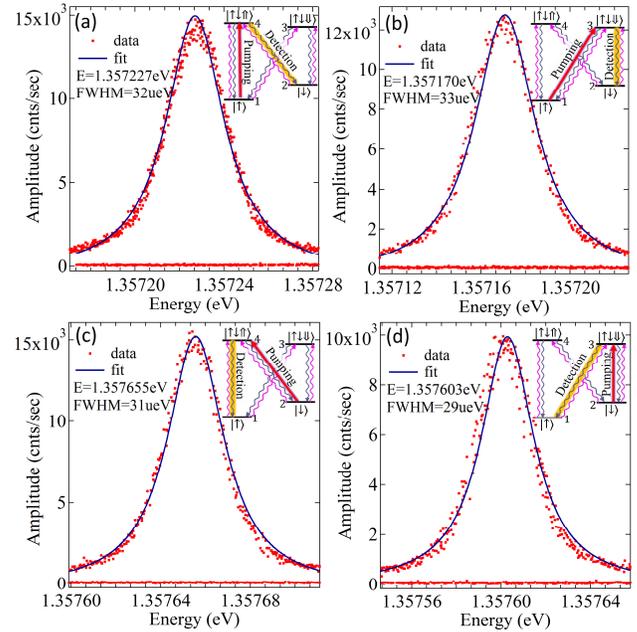

**Fig. S4.** Spin pumping of the nanowire quantum dot. (a) Pumping transition 1-4 while detecting photons coming from 4-2, (b) pumping 1-3 and detecting from 3-2, (c) pumping 2-4 and detecting 4-1 and (d) pumping 2-3 while detecting transition 3-1. When the randomization laser is turned off the counts drop to background level showing high fidelity spin initialization for all four transitions.

### References

1. B. J. Witek, R. W. Heeres, U. Perinetti, E. P. A. M. Bakkers, L. P. Kouwenhoven, and V. Zwiller, Measurement of the g-factor tensor in a quantum dot and disentanglement of exciton spins. *Phys. Rev. B* **84**, 195305 (2011).
2. K. G. Lagoudakis, P. L. McMahon, K. A. Fischer, S. Puri, K. Müller, D. Dalacu, P. J. Poole, M. E. Reimer, V. Zwiller, Y. Yamamoto and J. Vučković Initialization of a spin qubit in a site-controlled nanowire quantum dot. *New J. Phys.* **18**, 053024 (2016).